\documentclass[aps,prc,letterpaper,11pt,twoside,tightenlines,nofootinbib,showpacs,preprint]{revtex4}  %--> ORIGINAL

\usepackage{graphicx}
\usepackage{epsfig,float}
\usepackage[sort&compress]{natbib}
\usepackage{subfigure}
\usepackage{amsmath}
\usepackage{amsfonts}
\usepackage{cancel}
\usepackage{lmodern,dsfont}
\usepackage{amssymb}
\usepackage[colorlinks=true,urlcolor=blue,linktoc=page,linkcolor=blue,citecolor=blue]{hyperref}

\begin{document}
\arraycolsep1.5pt
\newcommand{\Ima}{\textrm{Im}}
\newcommand{\Rea}{\textrm{Re}}
\newcommand{\mev}{\textrm{ MeV}}
\newcommand{\gev}{\textrm{ GeV}}
\newcommand{\dtres}{d^{\hspace{0.1mm} 3}\hspace{-0.5mm}}
\newcommand{\rts}{ \sqrt s}
\newcommand{\non}{\nonumber \\[2mm]}
\newcommand{\eps}{\epsilon}
\newcommand{\half}{\frac{1}{2}}
\newcommand{\thalf}{\textstyle \frac{1}{2}}
\newcommand{\Nmass}{M_{N}} % mass of nucleon
\newcommand{\delmass}{M_{\Delta}} % mass of delta
\newcommand{\pimass}{\mu}  % mass of pion 
\newcommand{\rhomass}{m_\rho} % mass of rho
\newcommand{\piNN}{f}      % coupling of pi NN
\newcommand{\rhocoup}{g_\rho} % universal coupling to rho
\newcommand{\fpi}{f_\pi} % pion decay constant fpi
\newcommand{\f}{f} % pion decay constant fpi
\newcommand{\nucfld}{\psi_N} % nucleon field
\newcommand{\delfld}{\psi_\Delta} % delta field
\newcommand{\fpiNN}{f_{\pi N N}} % coupling of pi N N 
\newcommand{\fpiND}{f_{\pi N \Delta}} % coupling of pi N delta 
\newcommand{\GMquark}{G^M_{(q)}} % magnetic coupling for quark 
\newcommand{\vecpi}{\vec \pi}
\newcommand{\vectau}{\vec \tau}
\newcommand{\vecrho}{\vec \rho}
\newcommand{\delmu}{\partial_\mu}
\newcommand{\delMu}{\partial^\mu}
\newcommand{\nn}{\nonumber}
\newcommand{\bi}{\bibitem}
\newcommand{\vs}{\vspace{-0.20cm}}
\newcommand{\be}{\begin{equation}}
\newcommand{\ee}{\end{equation}}
\newcommand{\ba}{\begin{eqnarray}}
\newcommand{\ea}{\end{eqnarray}}
\newcommand{\ropi}{$\rho \rightarrow \pi^{0} \pi^{0}
\gamma$ }
\newcommand{\roeta}{$\rho \rightarrow \pi^{0} \eta
\gamma$ }
\newcommand{\omepi}{$\omega \rightarrow \pi^{0} \pi^{0}
\gamma$ }
\newcommand{\omeeta}{$\omega \rightarrow \pi^{0} \eta
\gamma$ }
\newcommand{\ul}{\underline}
\newcommand{\del}{\partial}
\newcommand{\rth}{\frac{1}{\sqrt{3}}}
\newcommand{\rsix}{\frac{1}{\sqrt{6}}}
\newcommand{\sq}{\sqrt}
\newcommand{\fr}{\frac}
\newcommand{\pr}{^\prime}
\newcommand{\ov}{\overline}
\newcommand{\Gm}{\Gamma}
\newcommand{\rw}{\rightarrow}
\newcommand{\rgl}{\rangle}
\newcommand{\De}{\Delta}
\newcommand{\Dp}{\Delta^+}
\newcommand{\Dm}{\Delta^-}
\newcommand{\Dz}{\Delta^0}
\newcommand{\Dpp}{\Delta^{++}}
\newcommand{\Sg}{\Sigma^*}
\newcommand{\Sp}{\Sigma^{*+}}
\newcommand{\Sm}{\Sigma^{*-}}
\newcommand{\Sz}{\Sigma^{*0}}
\newcommand{\X}{\Xi^*}
\newcommand{\Xm}{\Xi^{*-}}
\newcommand{\Xz}{\Xi^{*0}}
\newcommand{\Om}{\Omega}
\newcommand{\Omm}{\Omega^-}
\newcommand{\kp}{K^+}
\newcommand{\kz}{K^0}
\newcommand{\pip}{\pi^+}
\newcommand{\pim}{\pi^-}
\newcommand{\piz}{\pi^0}
\newcommand{\et}{\eta}
\newcommand{\kb}{\ov K}
\newcommand{\km}{K^-}
\newcommand{\kbz}{\ov K^0}
\newcommand{\ksb}{\ov {K^*}}

\newcommand{\dd}{\mathrm{d}}

\def\tstrut{\vrule height2.5ex depth0pt width0pt} % used in tables
\def\jtstrut{\vrule height5ex depth0pt width0pt} % used in tables

\title{\boldmath Constraints on a possible dibaryon from combined analysis of the $pn \to d \pi^+ \pi^-$ and $pn \to pn \pi^+ \pi^-$ cross sections}

\author{M. Albaladejo and E. Oset}
\affiliation{
Departamento de F\'{\i}sica Te\'orica and IFIC, Centro Mixto Universidad de 
Valencia-CSIC\\
Institutos de Investigaci\'on de Paterna, Aptdo. 22085, 46071 Valencia,
Spain
}

\date{\today}

\begin{abstract}
We use recent data that show a narrow peak around $\sqrt{s}=2.37\ \text{GeV}$ in the $pn \to d \pi^+ \pi^-$ cross section, with about double strength at the peak than in the analogous $pn \to d \pi^0 \pi^0$ reaction, and, assuming that it is due to the excitation of a dibaryon resonance, we evaluate the cross section for the $pn \to pn \pi^+ \pi^-$ reaction, with the final $pn$ unbound but with the same quantum numbers as the deuteron.  We use accurate techniques to determine the final state interaction in the case of the $pn$ forming a deuteron or a positive energy state, which allow us to get the $pn \to pn \pi^+ \pi^-$ cross section with $pn$ in $I=0$ and $S=1$, that turns out to be quite close or saturates the experimental $pn \to pn \pi^+ \pi^-$ total cross section around $\sqrt{s} = 2.37\ \text{GeV}$, depending on the angular momentum assumed. We then parametrize a background with different methods, and the sum of the resonant and background contributions is fitted to present data. The resulting cross section exceeds the experimental results in the region of the resonant peak, showing a problem in the dibaryon hypothesis. Yet, in view of the dispersion of present experimental data, and the scarce information around $\sqrt{s} = 2.37\ \text{GeV}$, a call is made for precise measurements of the $pn \to pn\pi^+\pi^-$ reaction around this energy, to further clarify this issue.

\end{abstract}
%\pacs{11.80.Gw --> ?, 12.38.Gc --> Lattice QCD Calc, 12.39.Fe --> Chiral Lagrangians, 13.75.Lb --> Meson-Meson interactions}
% 14.20.Gk --> baryon resonances ????
% Neutron-proton interactions, 13.75.Cs [NN interactions, in general], 13.85.-t
\maketitle

\section{Introduction}
\label{Intro} 

The $pn \to d \pi^0 \pi^0$ reaction has shown an intriguing feature since the cross section exhibits a very narrow peak around $\sqrt{s}= 2.37\ \text{GeV}$ of about $70\ \text{MeV}$ \cite{exp1,exp2}. The invariant mass distribution also shows a preference for the two pions having an invariant mass close to two pion masses. The quantum numbers of the reaction demand that the $\pi^0 \pi^0$ state is in isospin $I=0$ which in turn also calls for $L_\pi=0$ of the pair of pions to allow them to go together and have an invariant mass closer to the threshold of the two pions. The narrowness  of the signal has prompted the authors of \cite{exp1,exp2} to claim  that it is most probably due to the formation of a dibaryon resonance. The search for dibaryons has been a recurring subject, so far filled only with negative results. The confirmation of a dibaryon as being responsible for the reaction would undeniably be an important discovery. 

The first problem one faces in the dibaryon interpretation is that the $pn \to d \pi^0 \pi^0$ reaction is done as a fusion process. The $pn$ scattering is actually done using a deuteron target and arguments are given in \cite{exp1,exp2} to show that the proton in the deuteron is acting as a spectator. Hence, the reaction proceeds via the breakup of the target deuteron followed by the recombination of the deuteron from the neutron of the target and the proton of the beam, according to \cite{exp1,exp2}. One might think that it is easier to make the scattering on the original deuteron that is not broken and remains in the final state, and actually this is an unavoidable part of the mechanisms of the actual reaction, although the cuts made to demand that the proton of the initial deuteron acts as a spectator should make that mechanism subdominant. Yet, it is unclear what its unavoidable interference with the dominant mechanism could be. 

One can look for other possible reasons for the narrow peak. Assuming the fusion reaction to be responsible for it, if one has a long range $t$-channel mechanism for the production (imagine for instance $\Delta \Delta$ production mediated by pion exchange), the fusion reaction involves the deuteron wave function in momentum space in a way that could magnify certain kinematics. The fact is that although possible explanations could be given related to the the way the deuteron wave function enters the fusion reaction, no theoretical study along these lines has been done. The hypothesis of the dibaryon resonance then stands without a contradiction so far.

One step forward to show possible contradictions was given in Ref.~\cite{wilkin}. In this paper the authors take the experimental cross section of the  $pn \to d \pi^0 \pi^0$ reaction, assuming it to be formed by a resonance in the entrance channel, and relate it to the cross section of the $pn \to pn \pi^0 \pi^0$ reaction with the positive energy $pn$ system in the final state having the same quantum numbers as the deuteron, $I=0$, $S=1$. Using approximate techniques to take into account the final state interaction in the case of $d$ or $pn$ formation, the authors can determine the resonant cross section for  $pn \to pn \pi^0 \pi^0$ reaction, having a strength at the peak of about $0.4\ \text{mb}$.\footnote{The authors of \cite{wilkin} study the case where the angular momentum between the pair of pions and the deuteron is $L=0$ or $L=2$. Since $L=2$ is the case favored in \cite{exp1,exp2} we refer to the numbers of \cite{wilkin} for $L=2$.} To this cross section, one should add the background contribution from many other quantum numbers in the final state and compare with the total experimental cross section for the  $pn \to pn \pi^0 \pi^0$ reaction. Unfortunately there are no data for this reaction and the authors urged the experiment to be performed to clarify the issue. 
  
  In \cite{wilkin} the consistency with the inelasticity in the $NN$ cross section was also investigated and it was found that the inelastic $NN$ cross section based only on the resonant mechanism exceeded the experimental cross section of the $^3D_3$ partial wave obtained from the SAID analysis of \cite{arndt}, and was barely below the sum of the two partial waves $^3D_3 + ^3G_3$, which are possible with $J^P = 3^+$ in the entrance channel. In between, the normalization of the data used for the  $pn \to d \pi^0 \pi^0$ reaction has been reduced from $0.4\ \text{mb}$ at the peak to $0.27\ \text{mb}$ in \cite{pippim} and now the estimated $NN$ inelastic cross section from the  $pn \to pn \pi^0 \pi^0$ reaction with $pn$ having the deuteron quantum numbers alone would be well below the experiment.\footnote{C. Wilkin, private communication.}

  In between, some important experimental information has appeared from the measurement of the $pn \to d \pi^+ \pi^-$ cross section around the peak of the $pn \to d \pi^0 \pi^0$ one. Since the $\pi^+ \pi^-$ state has also $I=0$ component, it was certainly a puzzle that the narrow peak seen in the $pn \to d \pi^0 \pi^0$ cross section was not seen in the $pn \to d \pi^+ \pi^-$. Yet, the high precision data for this latter reaction measured in \cite{pippim} also show a clear and narrow peak exactly at the same position as the $pn \to d \pi^0 \pi^0$ reaction, with a strength at the peak about twice as big as for  $pn \to d \pi^0 \pi^0$, as demanded by isospin symmetry. 
  
  With this latter information it becomes most advisable to conduct the same test as in \cite{wilkin}, relating the $pn \to d \pi^+ \pi^-$ and $pn \to pn \pi^+ \pi^-$ cross sections, because now the predictions for the latter reaction can be compared with the experimental data \cite{dakhno,brunt,Tsuboyama:2000ub, Besliu:1986vd}. On the other hand, one can also use a more elaborate model to account for final state interaction than used in \cite{wilkin}, which was based on the use of the $pn$ scattering amplitude in the deuteron channel as being dominated by the deuteron pole. In the present paper we shall study the final state interaction using techniques developed in the chiral unitary approach \cite{kaiser,angels,ollerulf,carmenjuan,hyodo}. Using unitarity in coupled channels and the $N/D$ method it was found that it was justified to make an on shell factorization of the potential and the t-matrix in the Bethe Salpeter equation \cite{ollerulf,nsd} which renders this equation an algebraic one for each partial wave. Furthermore, we shall also use results from \cite{daniel,junko,francesca}, where the link between scattering amplitudes of the chiral unitary approach and wave functions in coordinate space is made and a meaning is found for the couplings of a bound state or a resonance to the interacting particles. Using these theoretical techniques and the recent data for the $pn \to d \pi^+ \pi^-$ reaction from \cite{pippim}, we shall see that the predicted cross section for the $pn \to pn \pi^+ \pi^-$ reaction with $pn$ with the deuteron quantum numbers, assuming it to be due to the dibaryon formation, nearly equals the experimental total cross section for this reaction at the energy where the $pn\to d\pi^+\pi^-$ cross section peaks. By adding a necessary background, which is obtained by a fit to the $pn \to pn\pi^+\pi^-$ data adding the resonance contribution and a background parametrized in different ways, the total cross section obtained exceeds the experimental one around the energy of the resonant peak. While this poses a problem to the dibaryon hypothesis, we also show that the data are scarce around $\sqrt{s} = 2.37\ \text{GeV}$, and we urge that precise experiments around this region are performed to clarify this issue. 
  
\section{Formalism}
\begin{figure}
\includegraphics[height=9cm,keepaspectratio]{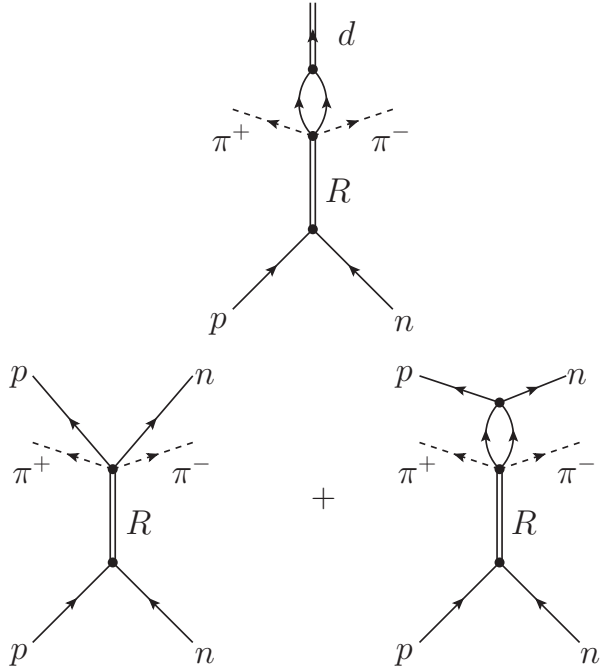}
\caption{Feynman diagrams leading to the formation of the deuteron in the $pn\to d\pi^+\pi^-$ reaction (top) and to scattering of $pn$ with positive energy (final state interaction) in the $pn \to pn \pi^+\pi^-$ reaction (bottom).\label{fig:diagrams}}
\end{figure}  

In Refs.~\cite{exp1,exp2} the dibaryon resonance ($R$) is assumed to have $J^P = 3^+$ with an orbital momentum $L=2$ between the $\pi^+\pi^-$ and the deuteron. In principle, $L=0$ could also be acommodated but analysis of angular momentum in \cite{exp1,exp2} favours $L=2$. As in Ref.~\cite{wilkin}, we shall study both cases.

In Fig.~\ref{fig:diagrams} we show the Feynman diagrams that lead to the formation of the deuteron ($d$) in the $pn \to d\pi^+\pi^-$ reaction or to the final state interaction (FSI) of the $pn$ pair in the $pn \to pn\pi^+\pi^-$ reaction. In the first case (upper diagram), two nucleons are produced from a bare $R \to pn\pi^+\pi^-$ vertex, which interact among themselves to form the deuteron. In the second case (lower diagrams), we have the bare $R \to pn\pi^+\pi^-$, corresponding to no interaction of the $pn$ pair, and the further interaction of this pair. Note that the interaction of the $pn$ pair is needed to form the deuteron.

Let $V_P$ be the bare vertex for $R\to pn \pi^+\pi^-$. We shall assume, as in \cite{wilkin}, that this corresponds to a short range process, as one expects from a compact object of a resonance. Following the idea of \cite{exp1,exp2} that this object is related to a double $\Delta$ state, the two $\Delta$'s would be bound by $93\ \text{MeV}$, which certainly makes the system rather compact. The transition matrix, $t_P$, from $pn$ to $\pi^+\pi^- d$ is then given by:\footnote{We neglect the small $D$-wave component of the deuteron.}
\begin{equation}\label{eq:tPd}
t_P^{(d)} = V_P G(M_d) g~,
\end{equation}
where $g$ is the coupling of the deuteron to $pn$ ($g^2$ is the residue of the $pn$ scattering matrix at the deuteron pole) and $G$ is the $pn$ propagator or loop function of the $p$ and $n$ propagators evaluated at the deuteron mass, $M_d$. The $G$ function is given, in the case of two nucleons, by \cite{angels}:
\begin{equation}
G(W) = i\int \frac{d^4 q}{(2\pi)^4} \left(\frac{M}{E(q)}\right)^2 \frac{1}{q^0 - E(q) + i\epsilon} \frac{1}{W-q^0-E(q)+i\epsilon}~,
\end{equation}
with $W$ the CM energy of the $d$ or the final $pn$ system, and $M$ the average nucleon mass. Upon Cauchy integration of the $q^0$ variable, $G(W)$ can be written as:
\begin{equation}\label{eq:Loop}
G(W) = \int \frac{d^3 q}{(2\pi)^3} \left(\frac{M}{E(q)}\right)^2 \frac{1}{W - 2E(q) + i\epsilon}~,
\end{equation}
which is conveniently regularized with a cut off $q_\text{max}$ by means of $\theta(q_\text{max} - \lvert \vec{q} \rvert)$. The $G$ function that appears here is the same that appears in the Bethe-Salpeter equation (BSE) (Lippmann-Schwinger with relativistic propagators), which in its on shell factorized form \cite{nsd, ollerulf} reads:
\begin{equation}\label{eq:BSE}
t = v + vGt~, \qquad t = \left( v^{-1} - G \right)^{-1}~,
\end{equation}
where $v$ is now the potential and $t$ the scattering matrix. Note that in the ordinary Lippmann-Schwinger equation $v$ and $t$ of the second term on the right hand side of the first form of Eq.~\eqref{eq:BSE} would go inside the integral of Eq.~\eqref{eq:Loop} with its half off shell form. The on shell factorization, where one neglects the left hand cut of the dispersion relation (which can be easily acommodated with a suitable small change in the cut off), allows to factorize $v$ and $t$ with their on shell value outside the integral, leading to Eq.~\eqref{eq:BSE}. One can actually perform a fit in the $pn$ data and determine $v$ and the cut off providing a good fit to experiment, but we shall not need it here.

The factorized form of the BSE is rather useful since it allows us to write:
\begin{equation}
t = v \left( 1 + Gt \right)~, \qquad 1 + Gt = \frac{t}{v}~.
\end{equation}
Let us call $t$ of Eq.~\eqref{eq:BSE} $t_{pn,pn}$ in what follows for clarity. If we go to the case of unbound $pn$ production in Fig.~\ref{fig:diagrams} (lower panel), the transition $t$-matrix will be given by:
\begin{equation}\label{eq:tPpn}
t_P^{(pn)} = V_P + V_P G(W) t_{pn,pn} = V_P \left( 1 + G(W) t_{pn,pn} \right)= V_P \frac{t_{pn,pn}}{v}~.
\end{equation}
Now it is useful to make use of the fact that the $t_{pn,pn}$ scattering matrix has a pole at the deuteron mass, where we have:
\begin{equation}
t_{pn,pn}(W) = \frac{g^2}{W - M_d}~,\quad W \simeq M_d~.
\end{equation}
Now, if we look at the expression of $t_{pn,pn}$ in Eq.~\eqref{eq:BSE}, we can see that if there is a pole at $W = M_d$ then:
\begin{equation}\label{eq:v_indep_s}
v^{-1} = G(M_d)~.
\end{equation}
Furthermore, upon use of L'H\^opital rule on the last form of Eq.~\eqref{eq:BSE}, we can see that:
\begin{equation}\label{eq:g2_def}
g^2 = \lim_{W \to M_d} \left( W - M_d \right) t_{pn,pn} = \left. \frac{1}{-\frac{d G}{dW}} \right\rvert_{W = M_d}~,
\end{equation}
assuming that $v$ is energy independent, which is a quite good approximation to obtain phase shifts by means of Eq.~\eqref{eq:BSE}.

By using Eq.~\eqref{eq:v_indep_s} we can now go back to Eq.~\eqref{eq:tPpn} and write:
\begin{equation}\label{eq:tPpn_2}
t_P^{(pn)} = V_P G(M_d) t_{pn,pn}~.
\end{equation}
Now, comparing Eqs.~\eqref{eq:tPpn_2} and \eqref{eq:tPd}, we find that:
\begin{equation}\label{eq:ratio_form_1}
\frac{ t_P^{(pn)} }{ t_P^{(d)} } = \frac{1}{g} t_{pn,pn}~.
\end{equation}
Hence, in order to relate the cross sections for $pn\pi^+\pi^-$ and $d\pi^+\pi^-$ production all we need is to know $g$ and $t_{pn,pn}$.

The value of $g$ can be easily obtained from Eq.~\eqref{eq:g2_def}, and one can show that, in the limit of small binding, it is independent of the value of $q_\text{max}$ \cite{daniel}. In fact, using the result obtained in Ref.~\cite{daniel} adapted to our Field Theory normalization, we find:
\begin{equation}\label{eq:gamma_def}
g^2 = \frac{4\pi W\gamma}{M^3}~, \gamma = \sqrt{2\mu B} = \sqrt{MB}~,
\end{equation}
where $\mu$ is the reduced mass of the $pn$ system, $\mu = M/2$. Above, $B$ is the binding energy of the deuteron. This result is actually well known since it was first discussed in Ref.~\cite{weinberg} in the context of the deuteron being a simple bound state of a proton and a neutron. This result has also been widely used to investigate if certain states qualify as dynamically generated states from an interaction potential or have a very different nature \cite{hanhart, baru}.

In our normalization, the $t_{pn,pn}$ scattering matrix is related to the one of Quantum Mechanics by:
\begin{equation}\label{eq:tpn_phases}
t_{pn,pn} = - \frac{2\pi W}{M^2} f^{\text{QM}} = - \frac{2\pi W}{M^2} \frac{1}{k\cot\delta(k) - ik}~,
\end{equation}
with $k$ the momentum of the particles in the CM frame. In order to be as model independent as possible, we take $t_{pn,pn}$ from experiment using the $pn$ phase shift of \cite{arndt}. As we can see, the ratio $\left\lvert t_P^{(pn)} / t_P^{(d)} \right\rvert^2$ has dimensions $\text{MeV}^{-3}$, as it should be, since in the evaluation of the cross sections there is an extra particle in the $pn \to pn \pi^+ \pi^-$ reaction and hence an extra $d^3 \vec{p} M/E(\vec{p})$ integration.

The cross section for the $pn \to d \pi^+\pi^-$ reaction is given by:
\begin{equation}\label{eq:sigma_d_def}
\sigma^{(d)}(s) = \frac{(2M)^2 2M_d}{2\left(s^2 - 4s M^2\right)^{1/2}}
\int\frac{\dd^3 p_1}{(2\pi)^3}\int\frac{\dd^3 p_2}{(2\pi)^3}\int\frac{\dd^3 p_d}{(2\pi)^3}
\frac{1}{2\omega_1}\frac{1}{2\omega_2}\frac{1}{2E_d} \left\lvert t_P^{(d)} \right\rvert^2 (2\pi)^4 \delta(P-p_1-p_2 - p_d)~,
\end{equation}
with $s$ the Mandelstam variable for the initial $pn$ system. Upon integration over the variables of the two pions, Eq.~\eqref{eq:sigma_d_def} gives:
\begin{equation}\label{eq:sigma_d_eva}
\sigma^{(d)}(s) = \frac{(2M)^2 2M_d}{2\left(s^2 - 4s M^2\right)^{1/2}}
\frac{\left\lvert t_P^{(d)} \right\rvert^2}{16\pi^3 \sqrt{s}} 
\int p_d\ \tilde{p}\ \dd M_\text{inv}~,
\end{equation}
where we have denoted by $M_\text{inv}$ the two-pion invariant mass, which is related to the deuteron energy by:
\begin{equation}
M_\text{inv}^2 = s + M_d^2 - 2\sqrt{s} E_d~.
\end{equation}
In Eq.~\eqref{eq:sigma_d_def}, $\tilde{p}$ is the pion momentum in the rest frame of the two pions,
\begin{equation}
\tilde{p} = \frac{\lambda^{1/2}\left( M_\text{inv}^2,m^2,m^2\right)}{2M_\text{inv}}~,
\end{equation}
being $m$ the pion mass, and $p_d$ is the deuteron momentum in the global CM frame,
\begin{equation}
p_d = \frac{\lambda^{1/2}\left(s,M_\text{inv}^2,M_d^2\right)}{2\sqrt{s}}~.
\end{equation}
In these equations, $\lambda(x,y,z)$ is the K\"ahlen or triangle function. Analogously, the cross section for the $pn \to pn \pi^+ \pi^-$ reaction is given by:
\begin{align}\label{eq:sigma_pn_def}
\sigma^{(pn)}(s) & = \frac{(2M)^4}{2\left(s^2-4sM^2\right)^{1/2}}
\int\frac{\dd^3 p_1}{(2\pi)^3}\int\frac{\dd^3 p_2}{(2\pi)^3}\int\frac{\dd^3 p_p}{(2\pi)^3}\int\frac{\dd^3 p_n}{(2\pi)^3} \nonumber \\
& \frac{1}{2E_p}\frac{1}{2E_n}\frac{1}{2\omega_1}\frac{1}{2\omega_2}
\left\lvert t_P^{(pn)} \right\rvert^2 (2\pi)^4 \delta(P - p_p - p_n - p_1 - p_2)~,
\end{align}
which upon integration of the two pion momenta gives:
\begin{align}\label{eq:sigma_pn_eva}
\sigma^{(pn)}(s) & = \frac{(2M)^4}{2\left(s^2-4sM^2\right)^{1/2}}\frac{1}{4(2\pi)^5}
\int \lvert\vec{p}_p\rvert \dd E_p \int \lvert\vec{p}_n\rvert \dd E_n \int_{-1}^{1} \dd \cos\theta \nonumber \\
& \left\lvert t_P^{(pn)} \right\rvert^2 \frac{\widetilde{p}}{M_\text{inv}} \theta(M_\text{inv} - 2m) \theta(\widetilde{M}_\text{inv} - 2M)~.
\end{align}
Here, $\widetilde{M}_\text{inv}$ is the $pn$ system invariant mass,
\begin{equation}
\widetilde{M}_\text{inv}^2 = 2M^2 + 2E_p E_n - 2\lvert\vec{p}_p\rvert \lvert\vec{p}_n\rvert \cos\theta~,
\end{equation}
with $p_p$ ($E_p$) and $p_n$ ($E_n$) referring the the momentum (energy) of the nucleons in the final $pn$ system. $M_\text{inv}$ is the two pion system invariant mass, now given by:
\begin{equation}
M_\text{inv}^2 = s + \widetilde{M}_\text{inv}^2 - 2 \sqrt{s} (E_p + E_n)~.
\end{equation}
In order to obtain the ratio of cross sections we need $\left\lvert t_P^{(pn)}/t_P^{(d)}\right\rvert^2$, which, according to Eqs.~\eqref{eq:ratio_form_1} and \eqref{eq:tpn_phases}, is given by:
\begin{equation}\label{eq:ratio_form_2}
\left\lvert \frac{t_P^{(pn)}}{t_P^{(d)}} \right\rvert^2 = \frac{4\pi^2s}{g^2 M^4} \left\lvert \frac{1}{k \cot \delta(k) - i k} \right\rvert^2 =
\frac{4\pi^2s}{g^2 M^4} \frac{\sin^2\delta(k)}{k^2}~,
\end{equation}
with $k$ given by:
\begin{equation}
k = \frac{\lambda^{1/2} \left( \widetilde{M}_\text{inv}^2,M^2,M^2 \right) }{2\widetilde{M}_\text{inv}}~.
\end{equation}

For the case $L=0$ we shall make use of Eqs.~\eqref{eq:sigma_d_eva} and \eqref{eq:sigma_pn_eva}. In the case $L=2$, however, we must take into account a factor $\vec{q}^4$ inside the integrals, where $\vec{q}$ is the relative momentum of the two-pion system with respect to the deuteron or the $pn$ system. This amounts to put the factor $\left\lvert \vec{p}_d \right\rvert^4$ for $\sigma^{(d)}$ in the integrand of Eq.~\eqref{eq:sigma_d_eva}, and $\left( \left\lvert \vec{p}_p \right\rvert^2 + \left\lvert \vec{p}_n \right\rvert^2 + 2\left\lvert \vec{p}_p \right\rvert \left\lvert \vec{p}_n \right\rvert \cos\theta \right)^2$ in the integrand of Eq.~\eqref{eq:sigma_pn_eva} for $\sigma^{(pn)}$.

It is interesting to compare our approach with that of Ref.~\cite{wilkin}. The results of this latter work are obtained by making the approximation of the scattering length for the amplitude $t_{pn,pn}$ of Eq.~\eqref{eq:ratio_form_1}, and demanding that it has a pole at the deuteron mass. Namely, $k\cot\delta$ in Eq.~\eqref{eq:ratio_form_2} is substituted by $-1/a$ with $1/a=\gamma$ of Eq.~\eqref{eq:gamma_def}.
\section{Results}
\begin{figure}[ht]
\includegraphics[width=0.47\textwidth,keepaspectratio]{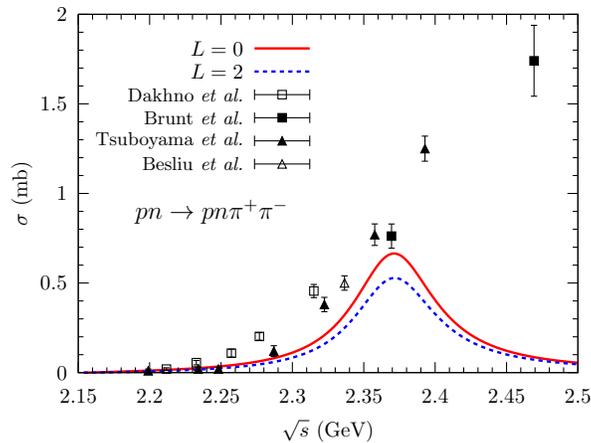}
\caption{Cross sections for the $pn\to pn\pi^+\pi^-$ reaction with $pn$ having the deuteron quantum numbers. Dashed line: prediction for $L=2$; solid line: prediction for $L=0$. Data are taken from Refs.~\cite{dakhno,brunt,Tsuboyama:2000ub, Besliu:1986vd}.\label{fig:cs_1}}
\end{figure}
Once the ratio $\sigma^{(pn)}(\sqrt{s})/\sigma^{(d)}(\sqrt{s})$ is calculated from Eqs.~\eqref{eq:sigma_d_eva} and \eqref{eq:sigma_pn_eva}, $\sigma^{(pn)}$ can be calculated from $\sigma^{(d)}(\sqrt{s})$. We parameterize the experimental cross section for $pn \to d\pi^+\pi^-$ in terms of a Breit-Wigner with mass $M_R = 2.37\ \text{GeV}$ and width $\Gamma = 70\ \text{MeV}$, with a peak $\sigma(\sqrt{s}=M_R) = 0.44\ \text{mb}$.\footnote{The value at the peak in Ref.~\cite{pippim} is $0.5\ \text{mb}$, but to this one must subtract the background from $I=1$ making also small corrections for different phase space for charged and neutral pions.} The cross sections that we obtain for $pn \to pn\pi^+\pi^-$ with the final $pn$ having the deuteron quantum numbers is shown in Fig.~\ref{fig:cs_1} for $L=0$ and $L=2$, together with the experimental data for the total cross section of $pn \to pn \pi^+\pi^-$ \cite{dakhno,brunt}. As we can see, for the case $L=0$ the predicted cross section is barely below the experimental point at $\sqrt{s} \simeq 2.37\ \text{GeV}$. For $L=2$ the cross section is about 25\% below the datum.

It is interesting to compare these results with those found in Ref.~\cite{wilkin} for the $pn \to pn\pi^0\pi^0$ reaction. The results are qualitatively similar, the cross section for $L=0$ being larger than that for $L=2$. The ratio of cross sections $\sigma^{(pn)}/\sigma^{(d)}$ is $0.9$ in Ref.~\cite{wilkin} for $L=2$, while in our case it is $1.2$ for the charged pions reactions, and it would be the same for the neutral pion reactions. The qualitative agreement is quite good in view of the simplifications done in Ref.~\cite{wilkin}, as quoted above.

\begin{figure}[ht]
\includegraphics[width=0.47\textwidth,keepaspectratio]{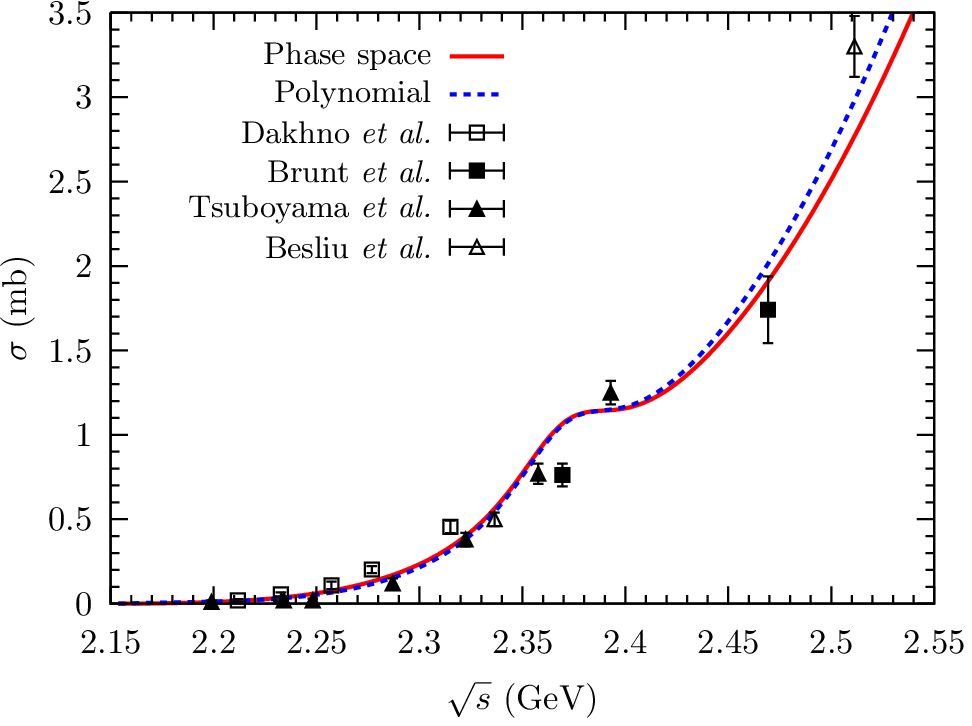}
\includegraphics[width=0.47\textwidth,keepaspectratio]{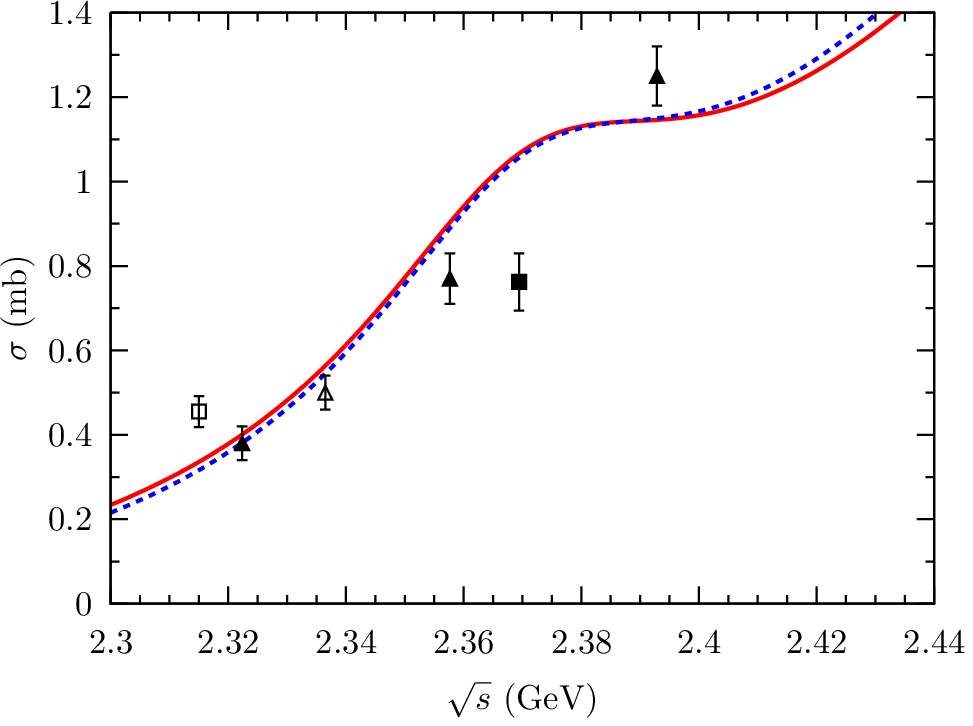}
\caption{Cross sections for $pn \to pn\pi^+\pi^-$ when a background is added to the resonant cross section. The solid line includes a background determined through phase space, while the dashed line includes a polynomial background. Data are taken from Refs.~\cite{dakhno,brunt,Tsuboyama:2000ub, Besliu:1986vd}. The left panel shows the whole range $2.15-2.55\ \text{GeV}$, whereas the right panel contains the energy range $\sqrt{s} = M_R \pm \Gamma$.\label{fig:cs_2}}
\end{figure}
The results of Fig.~\ref{fig:cs_1} indicate that if we add some background for the $pn \to pn \pi^+ \pi^-$ reaction from all other contributions where the $pn$ does not have the deuteron quantum numbers, or from the standard non-resonant mechanisms of two-pion production present in theoretical models \cite{luis, cao},\footnote{The model of Ref.~\cite{cao} does not sum amplitudes for the different diagrams but cross sections. Still, it provides fair integrated cross sections in the different channels.} one would get sizeable cross sections that might exceed the experimental cross sections at the peak of the resonant contribution. In order to show this, we have made some estimation of the background in two ways. In one case, the background is taken from phase space. This is easily obtained, up to a constant factor, from Eq.~\eqref{eq:sigma_pn_eva} setting $t_P^{(pn)}/t_P^{(d)}$ to unity in the integrand. In the second case, we parameterize the background by a polynomial in the variable $P = \sqrt{s} - 2M - 2m$, that is:
\begin{equation}\label{eq:poly_bg}
\sigma_\text{BG} = c_1 P + c_2 P^2 + c_3 P^3~,
\end{equation}
where $c_i$ are arbitrary constants. We perform a best fit to the data suming the background and the resonant cross sections. The results so obtained are shown in Fig.~\ref{fig:cs_2}, only for the case $L=2$. We can see that in both cases (phase space and polynomial backgrounds) the cross section around the resonant peak exceeds the experimental cross section. Yet, we also observe that there is some dispersion of the data and the information around the relevant region of $\sqrt{s} = 2.37\ \text{GeV}$ is scarce. In view of this, we can only encourage measurements of the $pn \to pn\pi^+\pi^-$ cross section with good resolution in the range $\sqrt{s} = 2.30$--$2.45\ \text{GeV}$ in order to clarify the situation and show wether there is or not a strong resonant peak.

\begin{figure}[ht]
\includegraphics[width=0.47\textwidth,keepaspectratio]{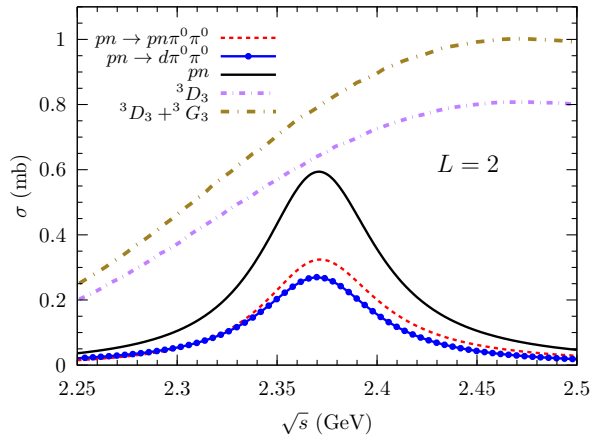}
\caption{Predicted cross sections for the $pn \to (d/pn)\pi^0\pi^0$ reactions. The blue solid line with circles represents a Breit-Wigner parameterization of the experimental cross section for $pn\to d\pi^0\pi^0$ of Ref.~\cite{pippim}, with $\sigma=0.27\ \text{mb}$ at the peak. Our prediction for the $pn \to pn\pi^0\pi^0$ reaction (having the final $pn$ the deuteron quantum numbers) is shown with the red dashed line. The black solid line results from the sum of both contributions. The upper dot-dashed lines represent the SAID analysis \cite{arndt} for the $NN$ inelastic cross section for the $^3D_3$ and the $^3D_3 + ^3G_3$ partial waves, with an isospin factor $1/6$.\label{fig:cs_3}}
\end{figure}
We also perform the same test as done in Ref.~\cite{wilkin} to compare the inelastic cross section for $NN$ scattering with $I=0$ with the cross section obtained for the $pn \to pn \pi^+\pi^-$ reaction. From isospin considerations the $I=0$, $J^P = 3^+$ contribution from $NN \to NN \pi\pi$ would be three times the cross section of $pn \to pn\pi^+\pi^-$ or six times that of $pn \to pn\pi^0\pi^0$, to which we should add the contribution from $pn \to d\pi^+\pi^-$ or $pn \to d\pi^0\pi^0$, respectively. In view of this, in Fig.~\ref{fig:cs_3}, we plot the cross section for $pn\to d \pi^0\pi^0$ and $pn\to pn \pi^0\pi^0$ and their sum as compared with one sixth of the inelastic cross sections for $NN$ in $I=0$, $J^P = 3^+$. Two possible initial partial waves can contribute in this case, the $^3D_3$ and the $^3 G_3$ waves. We show the inelastic cross sections corresponding to the SAID analysis \cite{arndt}. We see that the cross section evaluated from the two pion production is somewhat below the contribution from the $^3D_3$ and the sum of the $^3D_3$ and $^3G_3$ waves. It should be noted that, in spite of getting a larger ratio for $\sigma^{(pn)}/\sigma^{(d)}$ here than in \cite{wilkin} (1.2 versus 0.9) we still get a smaller $NN\to NN\pi\pi$ cross section. This is because the $0.4\ \text{mb}$ cross section used in \cite{wilkin} at the peak was changed to $0.27\ \text{mb}$ in \cite{pippim}. Thus, the argument used in \cite{wilkin} to put constraints on the resonance hypothesis is weakened when the new data are used.

\section{Conclusions}
We have carried out an exercise, taking into account the final state interaction of a $pn$ pair with the deuteron quantum numbers to produce a deuteron or a $pn$ pair with positive energy, in order to relate the cross sections of the $pn \to d \pi^+ \pi^-$ and $pn \to pn \pi^+ \pi^-$ reactions with the final $pn$ pair with the deuteron quantum numbers. The test was done assuming that the clear narrow peak seen in the $pn \to d \pi^+ \pi^-$ reaction was due to the formation of a dibaryon resonance in the $pn$ entrance channel, as suggested in the experimental papers of \cite{exp1,exp2,pippim}. We found that the cross section for this latter reaction was about 1.2 times bigger than for the  $pn \to d \pi^+ \pi^-$ reaction, assuming that one has $L=2$ angular momentum between the $pn$ and $\pi\pi$ pairs, as claimed in the experiments.  We compared this cross section with the total experimental one for the $pn \to pn \pi^+ \pi^-$ reaction and found that it was quite close to the experiment at the peak of the resonance. We then added a background from unavoidable standard nonresonant channels, which summed to the resonant contribution was fitted to present $pn \to pn \pi^+\pi^-$ data, and found that the total cross section exceeds the experimental one around $\sqrt{s}=2.37\ \text{GeV}$, where one has the peak of the $pn\to d\pi^+\pi^-$ cross section. We could admit uncertainties of about 20\% in the size of the $pn\to pn \pi^+\pi^-$ cross section (as also admitted in Ref.~\cite{wilkin}), but the fit with the resonant plus background contributions leads to results basically the same as in Fig.~\ref{fig:cs_2}. Hence, based on this scarce experimental information, this excess puts a problem to the hypothesis of the dibaryon resonance as being responsible for the narrow peak observed in the $pn \to d \pi^0 \pi^0$ and $pn \to d \pi^+ \pi^-$ reactions. 
In view of this, we can recommend two lines of research to further clarify this problem. The first one is to work on theoretical models for $NN \to NN \pi \pi$, and concretely for the $NN \to d \pi \pi$, reactions. For some mechanisms which involve a light particle (a pion) exchange in the $t$-channel, the consideration of the deuteron wave function is bound to constraint some distributions in the phase space that can repercute also in the cross section as a function of the energy. Some steps in this direction were done in \cite{luis2} in the study of the  $pn \to d \pi^0 \pi^0$ reaction using the model of \cite{luis}, but at much lower energies than those discussed here. The second line of suggested research is the measurement of the $pn \to pn \pi^+ \pi^-$ cross section with good resolution around the region of $\sqrt{s}= 2.37\ \text{GeV}$, where the peak of the cross sections in the deuteron production processes is seen. We have clearly shown that a large resonant cross section should develop as a consequence of the assumption of a resonant formation in the entrance channel. So far, the scarce data around this energy do not let us see the shape as a function of energy. A combination of efforts in those or other directions to clarify why such a narrow peak appears in the deuteron fusion reactions are much needed to see if the resonance hypothesis stands. We should note that many of the data points are very old, and present experimental facilities at COSY, HADES and other can provide measurements with far better accuracy. In view of this, we urge that experimental efforts are devoted to this task which, combined with the findings of the present paper, should help clarify this interesting problem.

\section*{Acknowledgments}  
We would like to acknowledge useful discussions with Colin Wilkin, Heinz Clement, Mikhail Bashkanov and Luis \'Alvarez-Ruso. This work is partly supported by the Spanish Ministerio de Economia y Competitividad and European FEDER funds under the contract number FIS2011-28853-C02-01 and FIS2011-28853-C02-02, and the Generalitat Valenciana in the program Prometeo, 2009/090. We acknowledge the support of the European Community-Research Infrastructure Integrating Activity
Study of Strongly Interacting Matter (acronym HadronPhysics3, Grant Agreement
n. 283286) under the Seventh Framework Programme of EU.

\end{document}